# Reconstructing Positions & Peculiar Velocities of Galaxy Clusters within 25000 km/sec: The Bulk Velocity

Enzo BRANCHINI[1,2], Manolis PLIONIS[1,3] & Dennis W. SCIAMA[1]

## ABSTRACT

Using a dynamical 3-D reconstruction procedure we estimate the peculiar velocities of $R \geq 0$ Abell/ACO galaxy clusters from their measured redshift within 25000 km/sec. The reconstruction algorithm relies on the linear gravitational instability hypothesis, assumes linear biasing and requires an input value of the cluster $\beta$-parameter ($\beta_c \equiv \Omega_\circ^{0.6}/b_c$), which we estimated in Branchini & Plionis (1995) to be $\beta_c \simeq 0.21$. The resulting cluster velocity field is dominated by a large scale streaming motion along the Perseus Pisces–Great Attractor base-line directed towards the Shapley concentration, in qualitative agreement with the galaxy velocity field on smaller scales. Fitting the predicted cluster peculiar velocities to a dipole term, in the local group frame and within a distance of $\sim 18000$ km/sec, we recover extremely well both the local group velocity and direction, in disagreement with the Lauer & Postman (1994) observation. However, we find a $\sim 6\%$ probability that their observed velocity field could be a realization of our corresponding one, if the latter is convolved with their large distance dependent errors. Our predicted cluster bulk velocity amplitude agrees well with that deduced by the POTENT and the da Costa et al. (1995) analyses of observed galaxy motions at $\sim 5000-6000$ km/sec; it decreases thereafter while at the Lauer & Postman limiting depth ($\sim 15000$ km/sec) its amplitude is $\sim 150$ km/sec, in comfortable agreement with most cosmological models.

[1]SISSA - International School for Advanced Studies, Via Beirut 2-4, I-34013 Trieste, Italy

[2]Present Address: Dept. of Physics, University of Durham. South Rd, DH1 3LE, Durham, U.K.

[3]National Observatory of Athens, Lofos Nimfon, Thesio 11810, Athens, Greece

## 1. Introduction

There is strong observational evidence for the existence of coherent large-scale galaxy flows in the local universe, extending from the Perseus-Pisces region on the one side to the Hydra-Centaurus/Great Attractor region on the other, pointing within $\lesssim 40°$ of the CMB dipole direction (see reviews and references in Dekel 1994 and Strauss & Willick 1995). These results, together with the large-scale coherence of the cluster gravitational acceleration (indicated by the fact that the differential cluster dipole in large equal-volume shells is aligned with the CMB dipole in each shell out to $\sim 16000$ km/sec; Plionis & Valdarnini 1991; Plionis, Coles & Catelan 1993; Plionis 1995), present a consistent picture in which the local group [hereafter LG] participates in a large-scale bulk motion induced by gravity, encompassing a volume of radius $\lesssim 15000$ km/sec.

This picture has recently been challenged by Lauer & Postman (1994) [hereafter LP94] who have extended the cosmic flow studies to very large scales using the brightest cluster galaxies as standard candles. They find that the LG motion with respect to the frame defined by the Abell/ACO clusters within 15000 km/sec moves in a direction $\sim 80°$ away from that of the CMB dipole, which then implies that, if the CMB dipole is a Doppler effect, the whole cluster frame is moving with respect to the CMB rest-frame with $\sim 700$ km/sec. Such a large velocity on such large scales is difficult to reconcile with galaxy bulk velocities on smaller scales, and with the current models of structure formation (cf. Strauss et al. 1995; Feldman & Watkins 1994). Furthermore it is difficult to understand why the LG peculiar acceleration, estimated from the cluster dipole, is well aligned with the mass (CMB) dipole (see Branchini & Plionis 1995, and references therein), while the observed LG peculiar velocity (as estimated by LP94), with respect to the same sample of clusters, is not.

The main aim of this Letter is to *predict* the Abell/ACO cluster velocity field and bulk flow, within the gravitational instability and linear biasing framework and compare it with determinations based on the *observed* peculiar velocities of galaxies (cf. POTENT) and of the LP94 clusters (see also Scaramella 1995 for complementary approach).

## 2. Method

We use the linear gravitational instability [GI hereafter] framework, linear biasing and an iterative technique, similar to that of Yahil et al. (1991), to reconstruct the Abell/ACO real cluster positions within 25000 km/sec, starting from their redshift space distribution, and thus obtain their peculiar velocity field. In the linear approximation the peculiar velocity $\mathbf{u}$ at position $\mathbf{r}$ is proportional to the gravitational acceleration (cf. Peebles 1980):

$$\mathbf{u}(\mathbf{r}) = \frac{\beta_c}{4\pi} \int \delta_c(\mathbf{r}) \frac{\mathbf{r}' - \mathbf{r}}{|\mathbf{r}' - \mathbf{r}|^3} \mathrm{d}^3\mathbf{r}' \qquad (1)$$

where $\beta_c \equiv \Omega_\circ^{0.6}/b_c$, $\Omega_\circ$ is the present value of the cosmological density parameter and $b_c$ is the usual linear bias parameter that relates the cluster and mass overdensities; $\delta_c = b_c \delta_m$. Note that the large relative separations of galaxy clusters causes them to sparsely trace the underlying density field; it is however this fact that make them ideal probes of the linear cosmic dynamics (cf. Bahcall et al. 1994), although non-linear effects could be present in high density regions, ie., up to scales of $\sim 1000$ km/sec (cf. Croft & Efstathiou 1994).

Our technique is a two step procedure whose complete description can be found in Branchini & Plionis (1995) [BP95 hereafter]; here we just recall the general idea.

- The first step consists in reconstructing the whole sky redshift-space cluster distribution with $cz \leq 25000$ km/sec. This is done by Montecarlo generating a population of synthetic objects outside the zone of avoidance (i.e. at $|b| > 20°$) and within 20000 km/s, whose spatial distribution accounts for galactic absorption, the radial selection function and density inhomogeneities between the Abell and ACO catalogues. Furthermore, the synthetic clusters are spatially correlated to the real ones according to the observed $\xi_{cc}(r)$. The object distribution within the zone of avoidance [ZoA hereafter] is then recovered by randomly cloning the cluster distribution within redshift–galactic longitude bins in the nearby latitude strips. The synthetic clusters within the [20000, 25000] km/sec shell are also distributed in order to minimize the known observational biases but spatially uncorrelated with real clusters, the latter to avoid introducing shot noise effects related to the exponential decrease of the real cluster number density. Finally, the density field beyond 25000 km/s is considered homogeneous and isotropic.

- The second step consists in applying to the whole–sky distribution of real + synthetic clusters reconstruction procedure that iteratively minimizes the reshift space distortions allowing to recover the 3-D cluster positions and peculiar velocities. This procedure assumes linear GI theory, linear biasing and requires an input $\beta_c$ parameter, which we take it to be $\beta_c = 0.21(\pm 0.03)$, obtained by comparing the amplitudes of the CMB and 3-D cluster dipoles (see BP95). To deal with possible non-linear effects on small scales we apply a top-hat smoothing of the forces.

An estimate of the intrinsic and possible systematic errors, introduced by modelling the observational biases (galactic absorption, radial selection and Abell/ACO homogenization scheme) which we call observational error, can be obtained by varying the model parameters in a plausible range and by generating different realizations of the cluster velocity field (for more details see BP95 and Plionis et al. *in preparation*). It turns out that for the input $\beta_c$ parameter used and for a smoothing radius of $10^3$ km/sec, the observational error distribution has a mean and standard deviation of $\langle \sigma_o \rangle \approx 80(\pm 60)$ km/sec. Similarly, the intrinsic uncertainty in the reconstruction procedure, $\sigma_I$, obtained as the scatter of the cluster velocities resulting from different Montecarlo realizations of the same model, is $\langle \sigma_I \rangle \approx 144(\pm 78)$ km/sec.

Another source of uncertainty is the approximate nature of the ZoA model and the possible systematic effects related to its increase with distance. In BP95 we implemented various schemes for filling the ZoA and found that the error on the reconstructed cluster positions was smaller than the intrinsic one and, more importantly, the resulting cluster dipole was nearly unaffected, which implies the stability of the bulk velocity measurement.

The parameters that affect mostly the resulting velocity field are the value of the input $\beta_c$ parameter which is required to start the iterative procedure and to a lesser extent the smoothing radius, $R_{sm}$, which affects however the velocities only in the high density regions. In BP95 we have verified that the choice of the input $\beta_c$ parameter does not bias the reconstruction procedure and that the so called "Kaiser effect" is unimportant in the present analysis.

## 3. The Local Group Velocity with respect to the Clusters

Similarly to LP94 we solve for the LG peculiar velocity with respect to the cluster frame, $\tilde{\mathbf{u}}_{LG}$, by minimizing:

$$\chi^2 = \sum_{i=1}^{N} \left\{ \frac{[cz_i - d_i] - \tilde{\mathbf{u}}_{LG} \cdot \vec{\mathbf{r}}_i}{\sigma_{T,i}} \right\}^2 , \qquad (2)$$

where $d_i$ is the reconstructed 3-D cluster distance in the LG frame, $\vec{\mathbf{r}}_i$ the unit position vector and $\sigma_{T,i}$ is the individual cluster total error computed by adding in quadrature $\sigma_{I,i}$, $\sigma_{o,i}$ and $\sigma_z(= 300$ km/sec) which represents the average uncertainty in $cz_i$. Note, that the $\chi^2$ significance measure of the dipole fit is ill-defined since the velocity errors are coupled. We have, however, investigated the stability of our solution to variations of the sample size (with $> 50\%$ reduction) and error-weighting and found that indeed it is very robust.

In table 1 we present the solution of eq.(2) for the $R_{sm} = 1$ and $2{\times}10^3$ km/sec cases and for three limiting radii of the volume used (boldface quantities refer to our standard case). It is important to note that out to the limiting depth of the LP94 sample ($r_{LP} \sim 15000$ km/sec) we find $|\tilde{\mathbf{u}}|_{LG}(r_{LP}) \approx 510 \pm 100$ km/sec with a misalignment angle with respect to the CMB dipole apex of only $\delta\theta_{cmb} \approx 5°$ while its asymptotic value ($\approx 635 \pm 70$ km/sec) is reached at $\sim 18000$ km/sec. This result, which assumes the GI hypothesis and agrees well with the observed CMB dipole, disagrees with the observed LG peculiar velocity, as measured by LP94, which has a similar amplitude but $\delta\theta_{cmb} \approx 80°$. This apparent discrepancy could be, among other things, due to the large distance dependent uncertainties of the LP94 velocities; in other words their velocity field could be a realization of an underlying field, represented by our reconstruction once convolved with the large distance dependent errors ($\sigma \approx 0.16\ r$).

Using Montecarlo simulations, in which we replaced each of our cluster 3-D distances, $r_i$, with $(1+\tilde{\epsilon}_i)r_i$ (where each $\tilde{\epsilon}_i$ is drawn from a Gaussian with $\mu = 0$ and $\sigma = 0.16$) and then fitting $\mathbf{u}_{LG}$ [via eq.(3)] for the resulting velocity fields, while using a $z^{-2}$ weighting (see LP94), we have found that in $\sim 6\%$ of the cases the derived $\tilde{\mathbf{u}}_{LG}$ was within $1\sigma$, in amplitude and direction, of that of LP94 (more details in Plionis et al *in preparation*).

Figure 1 shows $|\tilde{\mathbf{u}}|_{LG}$, the solution of eq.(2), computed within spheres of increasing radius together with the analogous quantity, $|\mathbf{u}|_g$, obtained from the gravitational acceleration induced by clusters (cluster dipole; see BP95) acting on the LG, after including a contribution of a $\sim 170$ km/sec Virgocentric infall and for $\beta_c = 0.21$ (error bars indicate 1 $\sigma_T$). There is a very good matching between the two quantities while their misalignment with the CMB dipole direction is $\lesssim 12°$ at the convergence depth. Although $\tilde{\mathbf{u}}_{LG}$ and $\mathbf{u}_g$ are not independent measures of the dipole, since they derive from the same underlying density field, their good matching constitutes a non–trivial demanding test of our reconstruction procedure, which we pass with success.

## 4. The Cluster Velocity Field and Bulk Flow

In figure 2 we present, for the $R_{sm} = 10^3$ km/sec case, the cluster peculiar velocity field in a 8000 km/sec wide slice projected onto the supergalactic plane, where most prominent superclusters lie (Hydra-Centaurus, Coma, Shapley, Perseus-Pisces, Ursa-Major and Grus-Indus). Open and filled dots refer to inflowing and outflowing objects, respectively, while the length of each vector is equal to 3 times the line of sight component of the peculiar velocity in the CMB frame. The small circle at the center represents the typical region spanned by dynamical analyses based on galaxy peculiar velocities (cf. Dekel 1994 and Strauss & Willick 1995). The most prominent feature is a large coherent motion in the general direction of the CMB dipole towards the Shapley Concentration $(X, Y) = (-13000, +9000)$ km/sec which does not have, however, a constant amplitude; it is small in the Perseus-Pisces region (X,Y)= (+8000, −4000) km/sec, then rises in the Great Attractor region (X,Y)= (−4000, +500) km/sec, while dropping on its backside (cf. Dressler & Faber 1991, Mathewson et al. 1994). Moreover, the bulk velocity rises again near the Shapley concentration where a strong back infall is apparent. Other features include the negligible peculiar velocities in the Coma region (cf. Courteau 1992) and the local infall in the Ursa-Major region, at (X,Y)=(7000,14000) km/sec.

We have measured the Abell/ACO cluster bulk velocity which is defined as the center of mass velocity of a specified region and is given by the integral of the cluster peculiar velocities $\mathbf{u}(\mathbf{x})$ over a selected volume specified by a selection function $\psi(\mathbf{x})$:

$$\mathbf{V}_{3D}^{bulk}(r_{max}) = \int_0^{r_{max}} \psi(\mathbf{x})\, \mathbf{u}(\mathbf{x})\, \mathrm{d}\mathbf{x} \qquad (3)$$

Assuming that clusters trace the mass, we can write eq.(3) for our discrete composite cluster sample as $\mathbf{V}_{3D}^{bulk} = \sum_i w_i \mathbf{u}_i / \sum_i w_i$, where the sums extend over both real and synthetic objects and $w_i$ is a weight that accounts for cluster masses and Abell/ACO sample differences (however, similar results are obtained even for $w_i = 1$). The discrete approximation need not to be equal to that of eq.(3) since it is a sum over the observed clusters which are inhomogeneously distributed. In our case, however, this sum extends over the whole real+synthetic cluster distribution and not only over the real objects. Moreover, we have found consistent estimates of $\mathbf{V}_{3D}^{bulk}$ using the cluster peculiar velocities either at the positions of the clusters or at a regular grid with grid size of 2000 km/sec.

The cluster bulk velocity can be also defined as the residual velocity of the whole cluster frame. Therefore an alternative estimator of the bulk velocity, which utilizes only real clusters and only the line of sight component of their peculiar velocities, which was also used by LP94, is:

$$\mathbf{V}_{1D}^{bulk}(r_{max}) = \mathbf{C} - \tilde{\mathbf{u}}_{LG}(r_{max}) \qquad (4)$$

where $\mathbf{C}$ is the CMB dipole vector and $\tilde{\mathbf{u}}_{LG}(r_{max})$ is given by eq.(2). Note that our reconstruction procedure assumes that the mass fluctuations responsible for the cluster peculiar motions are contained within the sample considered, imposing the bulk flow to vanish beyond 25000 km/sec. We have verified, varying the limiting sample depth and using an alternative reconstruction scheme that increases the amount of clustering in the 20000 – 25000 km/sec range (see BP95), that this constraint does not appreciably affect the bulk flow amplitude within 20000 km/sec.

In table 2 we present the results of eq.(4) as a function of sample limiting depth for two different smoothing radii. Note that the bulk flow at $r_{LP}$ has an amplitude of $\sim 150$ km/sec in comfortable agreement with currently accepted cosmological models. In figure 3 we plot, for $R_{sm} = 10^3$ km/sec, both estimators $\mathbf{V}_{3D}^{bulk}$ and $\mathbf{V}_{1D}^{bulk}$ as filled dots and starred symbols respectively, but now as a function of the *effective* depth ($r_{eff} \approx 3/4 r_{max}$). Note that for the $V_{1D}$ estimator we have used $r_{max} \geq 10000$ km/sec in order to have sufficient data for the $\chi^2$ minimization to be stable. Error bars represent 1 $\sigma$ total uncertainties. Although the two estimators are not independent, they are however based on a different set of clusters and velocities and they provide a consistent estimate of the bulk velocity. We also plot as open dots the bulk velocity obtained by the POTENT reconstruction of the Mark III velocity field (Dekel 1994) and as starred symbols the recent da Costa et al. (1995) bulk velocity. At the region where both, the galaxy and cluster bulk velocity estimates overlap (at $r_{eff} \sim 4000-6000$ km/sec ), the different bulk flow amplitudes appear to be in very good agreement with each other. Note that our bulk flow determination is mostly unaffected by the window 'shrinkage' effect (Kaiser 1988), since the cluster selection function has a value $\sim$ 1 up to $\sim$20000 km/sec and does not suffer from the use of only line of sight peculiar velocities (Regös & Szalay 1989) since the $\mathbf{V}_{3D}^{bulk}$ estimator is based on the full 3D velocity field.

## 5. Conclusions

An iterative reconstruction procedure, based on linear GI theory and linear biasing, has been used to derive the velocity field traced by Abell/ACO clusters within 25000 km/sec. Our main results are:

1. The predicted cluster velocity field, in the LG frame, is such that it reflects the whole LG motion with respect to the CMB, in apparent disagreement with the LP94 velocity data. Our derived LG velocity is well aligned with the CMB dipole and it reaches its asymptotic value at $\sim$ 18000 km/sec in agreement with the locally derived cluster dipole. We have found, a small but non-negligible probability ($\sim$ 6%) that the LP94 observed velocity field could be consistent with our predicted one; their apparent disagreement being possibly due to the convolution of the underlying velocity field with their large distance dependent errors.
2. The main features of the observed velocity field probed by galaxies (cf. Strauss & Willick 1995) are reproduced also by our predicted cluster velocity field. There is an evident extension of our cluster bulk flow out to $\sim$ 15000 km/sec, where a back-infall to the Shapley concentration is evident. The derived bulk flow velocity has an amplitude in very good agreement with that of POTENT and da Costa et al. (1995) at $\sim$ 5000 km/sec, it decreases thereafter while pointing towards the CMB dipole direction. Our predicted bulk velocity at $r_{max} \sim$ 10000 and 15000 km/sec is $\sim$300 and $\sim$150 km/sec, respectively, consistent with most theories of structure formation.

**Acknowledgements**

We thank the referee, M. Strauss, for many useful comments and criticisms. We also thank for a discussion F. Mardirossian and M. Mezzetti. MP acknowledges receipt of an EEC *Human Capital and Mobility Fellowship*. EB and DWS are grateful to Ministero dell'Universitá e della Ricerca Scientifica e Tecnologica for financial support.

## Figure Captions

**Figure 1:** Comparison between the LG peculiar acceleration, $|\mathbf{u}|_g$, as derived by the cluster dipole and the LG peculiar velocity, $|\tilde{\mathbf{u}}|_{LG}$, derived from eq.(2) for the $R_{sm} = 10^3$ km/sec case.

**Figure 2:** The cluster velocity field of a 8000 km/sec strip projected onto the supergalactic plane for the $R_{sm} = 10^3$ km/sec case. Open dots are inflowing clusters while filled dots are outflowing ones. The velocity vectors have been multiplied by a factor 3 to enhance their visibility.

**Figure 3:** The cluster bulk velocity as estimated by the two methods described in the text together with the galaxy based POTENT and the da Costa et al. (1995) values.

Table 1: Local Group motion with respect to the cluster frame within distance $r_{max}$ and for 2 different smoothing radii (both in $h^{-1}$ Mpc).

| $r_{max}$ | $R_{sm}$ | $\|\tilde{\mathbf{u}}\|_{LG}$ (km/sec) | $\tilde{l}$ | $\tilde{b}$ | $\delta\theta_{cmb}$ | $\chi^2/d.f.$ |
|---|---|---|---|---|---|---|
| 100 | **10** | **342± 102** | **260°± 38°** | **27°± 23°** | **16°** | **38/39** |
| | 20 | 316±80 | 276° ± 24° | 19° ± 16° | 11° | 26/38 |
| 150 | **10** | **512± 96** | **276°± 16°** | **27°± 11°** | **4°** | **202/116** |
| | 20 | 441±68 | 279° ± 12° | 28° ± 8° | 3° | 104/165 |
| 180 | **10** | **635± 70** | **278°± 9°** | **20°± 6°** | **11°** | **341/197** |
| | 20 | 570±55 | 278° ± 7° | 20° ± 5° | 10° | 193/197 |

Table 2: Residual bulk motion of cluster frame within distance $r_{max}$.

| $r_{max}/r_{eff}$ | $R_{sm}$ | $V_{1D}^{bulk}$ (km/sec) | $l$ | $b$ | $\delta\theta_{cmb}$ |
|---|---|---|---|---|---|
| 100/75 | **10** | **305** | **297°** | **31°** | **18°** |
| | 20 | 316 | 278° | 41° | 11° |
| 150/113 | **10** | **115** | **284°** | **45°** | **17°** |
| | 20 | 180 | 273° | 36° | 7° |

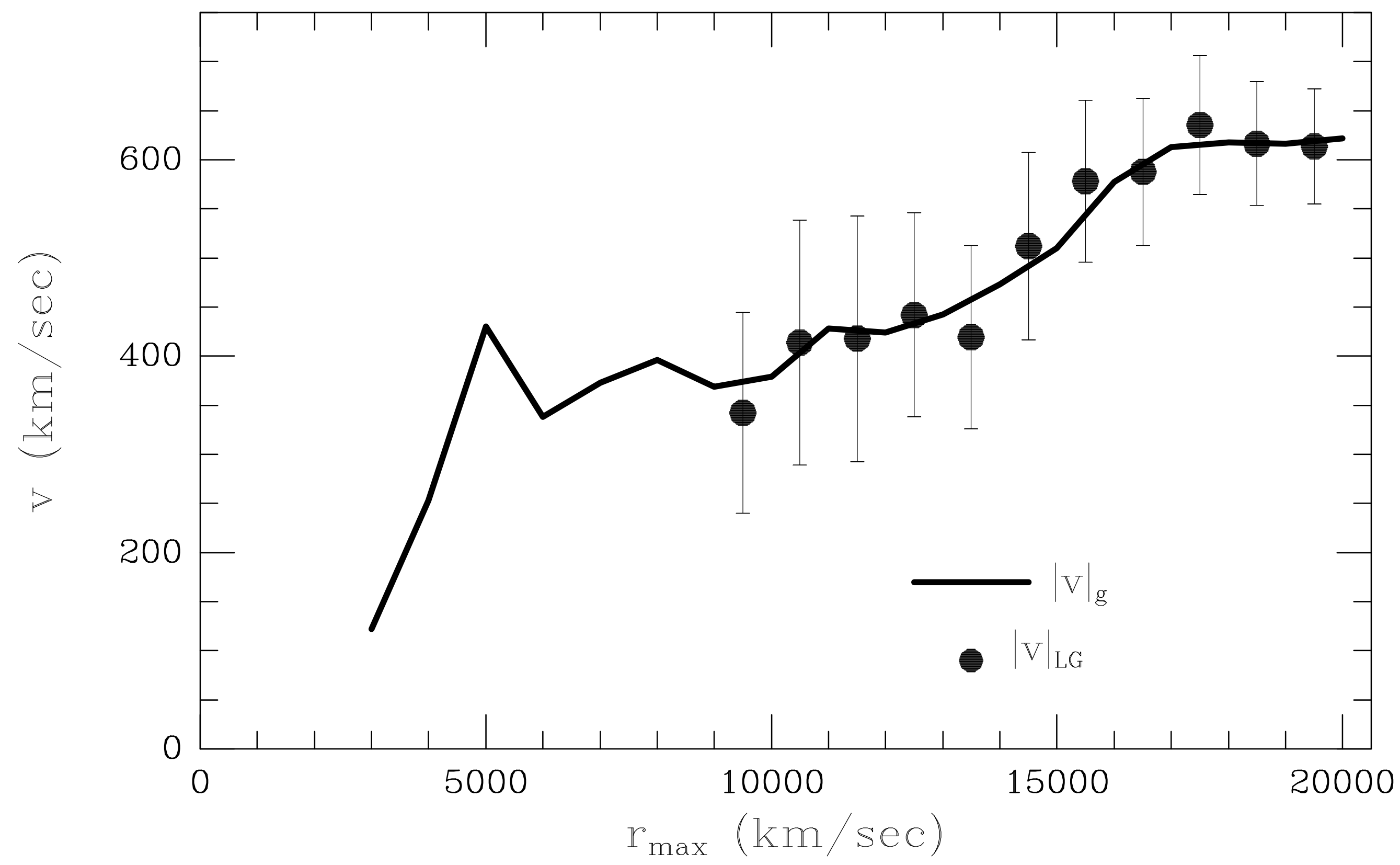

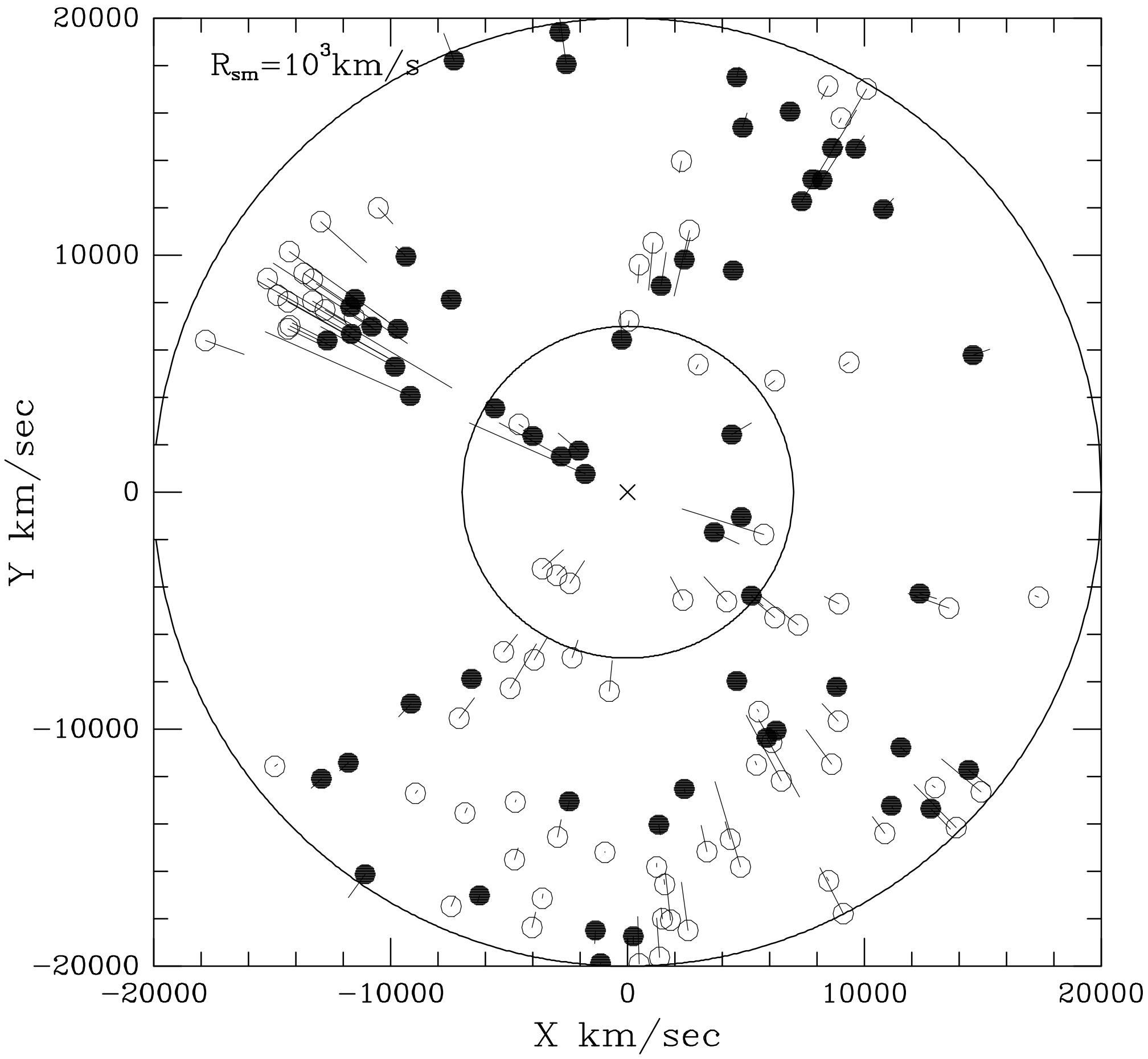

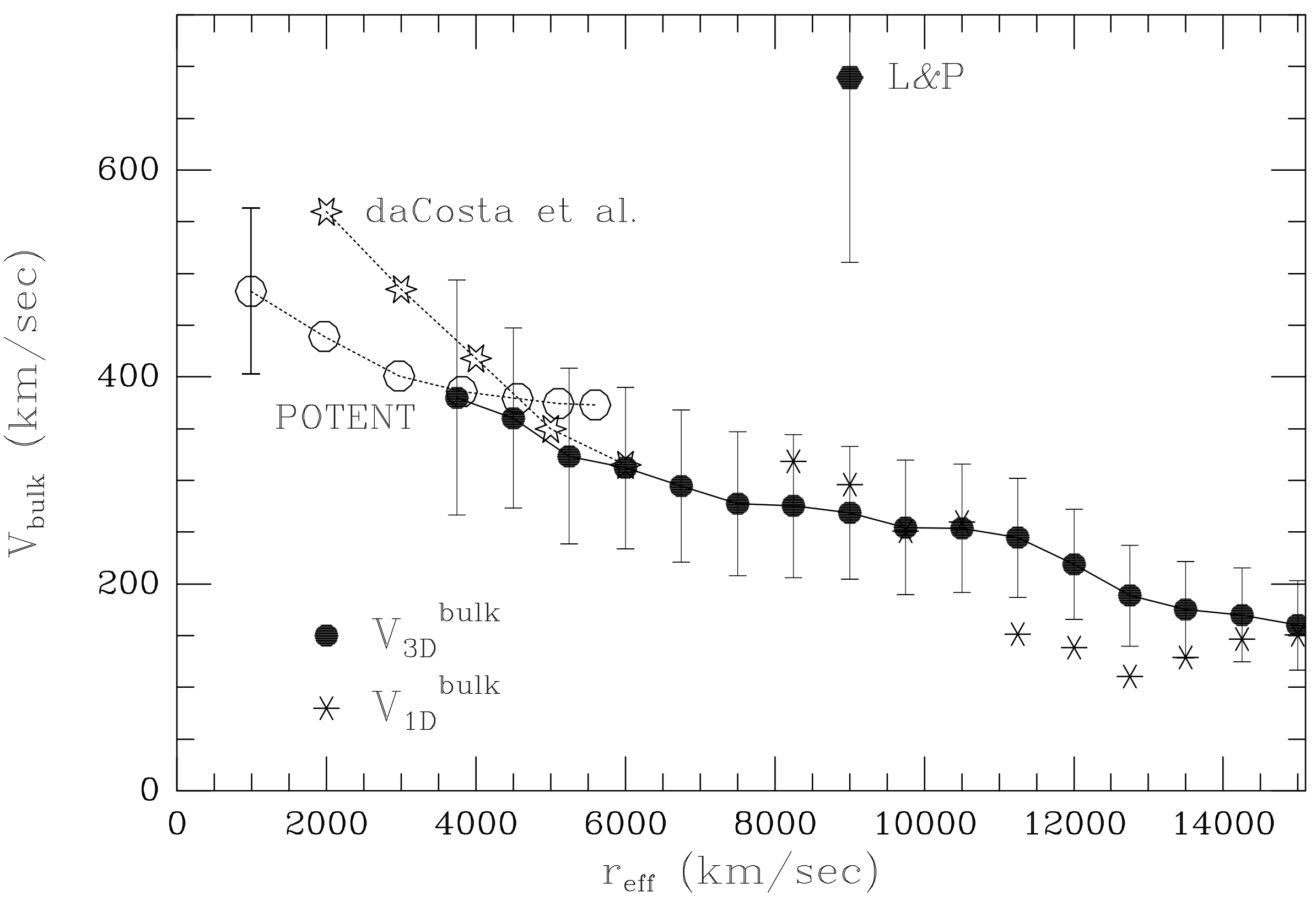